\documentclass[manuscript, screen, sigconf]{acmart}

\AtBeginDocument{%
  }

\setcopyright{rightsretained}
\copyrightyear{2025}
\acmYear{2025}
\acmDOI{XXXXXXX.XXXXXXX}

\acmConference[Conference acronym 'XX]{Make sure to enter the correct
  conference title from your rights confirmation emai}{June 03--05,
  2025}{Woodstock, NY}
\acmISBN{978-1-4503-XXXX-X/18/06}

\usepackage{longtable}
\usepackage[flushleft]{threeparttable}
\usepackage[export]{adjustbox} 
\usepackage{multirow}
\usepackage{pdflscape}
\usepackage{subcaption}
\usepackage{caption}
\usepackage{array}
\usepackage{colortbl} 

\definecolor{editCol}{rgb}{0.0, 0.0, 0.0}
\newcommand{\edit}[1]{{\textcolor{editCol}{#1}}}

\usepackage{comment}
\usepackage{balance}
\usepackage{booktabs} 

\begin{document}

\title[Teens Need to Be Educated on the Dangers]{`Teens Need to Be Educated on the Dangers’: Digital Access, Online Risks, and Safety Practices Among Nigerian Adolescents}


\author{Munachimso B. Oguine}
\orcid{}
\affiliation{%
\department{Computer Science}
  \institution{National Open University of Nigeria}
  \city{Abuja}
  \state{}
  \country{Nigeria}}
\email{munachioguine@gmail.com}

\author{Ozioma C. Oguine}
\orcid{}
\affiliation{%
\department{Computer Science and Engineering}
  \institution{University of Notre Dame}
  \city{Notre Dame}
  \state{Indiana}
  \country{USA}}
\email{ooguine@nd.edu}

\author{Karla Badillo-Urquiola}
\orcid{}
\affiliation{%
\department{Computer Science and Engineering}
  \institution{University of Notre Dame}
  \city{Notre Dame}
  \state{Indiana}
  \country{USA}}
\email{kbadillou@nd.edu}

\author{Oluwasogo Adekunle Okunade}
\orcid{}
\affiliation{%
\department{Computer Science}
  \institution{National Open University of Nigeria}
  \city{Abuja}
  \state{}
  \country{Nigeria}}
\email{aokunade@noun.edu.ng}

\renewcommand{\shortauthors}{Munachimso Oguine et al.}

\begin{abstract}
Adolescents increasingly rely on online technologies to explore their identities, form social connections, and access information and entertainment. However, their growing digital engagement exposes them to significant online risks, particularly in underrepresented contexts like West Africa. This study investigates the online experiences of 409 secondary school adolescents in Nigeria’s Federal Capital Territory (FCT), focusing on their access to technology, exposure to risks, coping strategies, key stakeholders influencing their online interactions, and recommendations for improving online safety. Using self-administered surveys, we found that while most adolescents reported moderate access to online technology and connectivity, those who encountered risks frequently reported exposure to inappropriate content and online scams. Blocking and reporting tools were the most commonly used strategies, though some adolescents responded with inaction due to limited resources or awareness. Parents emerged as the primary support network, though monitoring practices and communication varied widely. \edit{Guided by Protection Motivation Theory (PMT), our analysis interprets adolescents’ online safety behaviors as shaped by both their threat perceptions and their confidence in available coping strategies}. A thematic analysis of their recommendations highlights the need for greater awareness and education, parental mediation, enhanced safety tools, stricter age restrictions, improved content moderation, government accountability, and resilience-building initiatives. Our findings underscore the importance of culturally and contextually relevant interventions to empower adolescents in navigating the digital world, with implications for parents, educators, designers, and policymakers.

\end{abstract}

\ccsdesc[500]{Human-centered computing~Empirical studies in HCI}
\ccsdesc[500]{Social and professional topics~Adolescents}
\keywords{Nigerian Teens; Adolescents online safety; Global South; Culture; HCI4D; Youth Online Safety; Parental mediation}

\maketitle

\section{Introduction}
Adolescence is a pivotal developmental stage characterized by significant cognitive, emotional, and behavioral changes that shape youth self-awareness, decision-making, and ability to regulate their actions both offline and online. Online technologies have become integral to this phase of adolescnece, providing avenues for them to explore their identities \cite{sexting}, form social connections \cite{sexting}, and seek support \cite{oguine2024internet}, information \cite{genz_information}, and entertainment \cite{wanjiru2010study, namibia}. As digital natives, adolescents increasingly rely on online platforms to navigate various aspects of their lives. Sub-Saharan Africa, home to the world’s largest youth population with over 200 million youths \cite{africa_youth_stats}, is witnessing growing internet connectivity. According to the 2024 African Union Child Online Safety and Empowerment Policy \cite{Africa_policy}, 40\% of youth aged 15–24 years in Africa now have access to the internet. While this connectivity opens new opportunities, it also heightens teens’ exposure to significant online risks \cite{Africa_policy}.

Research on technology adoption among adolescents highlights its double-edged nature, offering substantial benefits while exacerbating vulnerabilities, especially for youth from marginalized backgrounds \cite{marginalized, Hendricks2020Apr, Oguine2023Feb}. African scholarship has explored how young users leverage online technologies while navigating the risks they encounter. For instance, Maoneke et al. \cite{namibia} examined ICT use among teens in Namibia, revealing exposure to cyberbullying, sexual risks, and privacy concerns while forming online connections. Similarly, studies in Kenya have documented adolescents’ experiences with online scams, stalking, and exposure to inappropriate content\cite{Oloo2022Apr}. However, much of this research has centered on regions outside West Africa, leaving gaps in understanding the specific experiences of adolescents in this area. Notably, there is a lack of empirical studies capturing self-reported accounts of Nigerian adolescents’ online experiences.

This study aims to address these gaps by extending current knowledge through a detailed exploration of the online experiences of secondary school adolescents in Nigeria by centering their perspectives. Hence, we propose the following research questions:
\edit{ 
\begin{itemize}
    \item \textbf{RQ1:} \textit{How do secondary school adolescents in Nigeria access and engage with online technologies, and what factors influence these interactions?}
    \item \textbf{RQ2:} \textit{What online risks do Nigerian secondary school adolescents face, and what strategies or literacy skills do they use to manage these risks?}
    \item \textbf{RQ3:} \textit{What roles do stakeholders and adolescents themselves play in shaping and promoting their online safety?}
\end{itemize}
}
To address these research questions, we conducted a self-administered survey with 409 secondary school adolescents in the Federal Capital Territory (FCT), Nigeria. The survey was designed to comprehensively capture adolescents' online experiences, focusing on their access to digital technologies, exposure to online risks, safety strategies, and the broader social ecology of online safety.
To address \textbf{(RQ1)}, participants provided detailed information about how they access and engage with online technologies. The findings revealed a moderate level of digital connectivity, as 59\% of our respondents reported regular access to the internet, several times a day. However, patterns of use were shaped by infrastructural constraints, such as limited device availability and inconsistent internet access. These constraints influenced the frequency, purpose, and quality of adolescents' online engagements. \textbf{(RQ2)} explored the online risks adolescents encountered online and the strategies or literacy skills they employed to manage them. While a majority reported limited exposure to serious online threats, a notable proportion of participants identified inappropriate content and online scams as the most common risks. Adolescents reported using a range of coping strategies, including blocking or reporting harmful content and seeking guidance from trusted adults. However, some participants indicated inaction when faced with online risks, often due to limited awareness or online safety education. Disparities in digital literacy were evident: while some adolescents reported having received online safety training, many others had not, or were unsure whether they had.

To address \textbf{(RQ3)}, our survey examined the roles of various stakeholders in shaping adolescents’ online safety. Parents emerged as the primary influencers, offering both monitoring and guidance. Other potential stakeholders, such as teachers, peers, and community figures, were inconsistently involved, reflecting a reliance on immediate family rather than extended social circle. To deepen our understanding of youth agency, participants were invited to share their own recommendations for improving online safety. Among the 268 responses to this open-ended question, thematic analysis revealed seven key suggestions. Adolescents called for more targeted awareness and education campaigns, emphasizing the need for relatable and accessible information about online risks. Parental mediation was frequently mentioned, with appeals for more active and informed supervision. Participants also advocated for improved safety and privacy tools, such as better blocking and reporting features, and stricter age restrictions. Content moderation especially combining algorithmic detection with human oversight, was seen as essential for reducing harmful exposure. Additionally, adolescents emphasized the need for government accountability, including stronger regulations and penalties for online offenders. Finally, resilience building emerged as a critical theme, with adolescents recommending interventions that equip them with critical thinking and emotional coping skills to navigate digital spaces more confidently. Our study makes several contributions to the fields of Human-Computer Interaction (HCI) and adolescent online safety:
\begin{itemize}
    \item \textbf{Empirical Insights:} By centering the voices of Nigerian adolescents', our study provides a detailed understanding of their online experiences, risks, and coping strategies, offering valuable data for both regional and global online safety research.
    \item \textbf{Focus on an Underrepresented Context:} We address a critical gap in literature by exploring the online safety experiences of adolescents in West Africa, specifically Nigeria, a region with unique cultural, technological, and socio-economic dynamics.
    \item \textbf{Actionable Recommendations:} Our findings offer practical guidance for parents, educators, designers, and policymakers in crafting educational initiatives, online features, and policies that are more inclusive, context-aware, and aligned with the needs of Nigerian adolescents.
\end{itemize}
In doing so, this research contributes to the AfriCHI community and the broader global discourse on adolescent online safety, emphasizing the need for culturally and contextually relevant interventions that empower adolescents to navigate the digital world safely and confidently.

\section{Background}
This section provides an overview of existing research on adolescent online safety, beginning with a global perspective and then narrowing the focus to studies conducted in Africa, particularly Nigeria.

\subsection{Adolescents Online Safety Research}
The rapid adoption of online technologies, including social media, gaming platforms, and educational tools, has fundamentally reshaped how adolescents interact with the digital world \cite{freed2023understanding}. Young users increasingly rely on these platforms to access information and educational resources \cite{eynon2012understanding}, engage in self-expression and community building \cite{knowles2025role}, seek social support \cite{oguine2024internet, ahola2018internet}, and connect with others \cite{razi_sext}, making them an integral part of their daily lives. While these technologies offer significant benefits, they also expose adolescents to various online risks, many of which have been widely documented within the SIGCHI research community \cite{akter2023takes, katie_resilienvce, freed2023understanding} and media reports \cite{mainstream_media}. Key concerns include cyberbullying \cite{freed2023understanding}, online sexual risks \cite{razi_sext}, privacy and data security threats \cite{priya-privacy}, and exposure to harmful content \cite{oguine2024internet}, all of which can have serious psychological and emotional consequences \cite{mchugh2017most}. For instance, Razi et al. \cite{razi_sext} found that teens aged 12–17 were highly vulnerable to sexting and online predation, particularly when engaging with strangers online. Similarly, Freed et al. \cite{freed2023understanding} highlighted the prevalence of online harassment, where youth were frequently targeted by peers, intimate partners, acquaintances, and strangers, further compounding their digital vulnerabilities.

To address these challenges, researchers have proposed various strategies to enhance adolescent online safety. These strategies range from parental monitoring and control \cite{parental_control} and active parental mediation \cite{Wisniewski2024Dec} to community engagement \cite{mamtaj_village} and promoting teen-autonomy \cite{katie_resilienvce}. For example, Wisniewski et al. \cite{wisniewski_moral} investigated the effects of restrictive versus active parental mediation, finding that excessive control over youth online activities often created tensions between parents and adolescents, with adolescents expressing concerns about privacy invasion. Recent research advocates for a developmental approach to online safety, recommending greater parental involvement for younger users, while emphasizing teen-autonomy and active mediation for older adolescents \cite{katie_resilienvce}. Agha et al. \cite{zainab_nudges} exemplified this approach by conducting co-design sessions with youth aged 13–18, exploring how nudging strategies can help teens make more informed decisions while using online technologies. Our research seeks to extend this body of work by examining the online safety experiences of Nigerian adolescents, addressing critical gaps in knowledge, and identifying culturally relevant interventions tailored to their needs. By capturing the lived experiences of Nigerian adolescents, this study aims to ensure that online safety strategies are not only contextually appropriate but also effective in protecting youth in digital spaces.

\subsection{African Research on Adolescent Online Safety}
Despite growing research on adolescent online safety, most studies remain Western-centric \cite{pinter2017adolescent}, often overlooking the socio-economic, cultural, and infrastructural realities of adolescents in the Global South. While online risks affect youth globally, the experiences of youth in low- and middle-income countries (LMICs) remain understudied \cite{Oguine2025Apr, Quayyum2024Jun, Livingstone2017, wilkinson2022many}, despite evidence suggesting that these regions face distinct challenges requiring context-specific interventions. Since the COVID-19 pandemic, African societies have witnessed a significant increase in technology adoption among young users \cite{Chipangura2022Dec}. Given both the opportunities and vulnerabilities that arise from increased digital engagement, several scholars \cite{namibia, Porter2020Jan, cyberbullet} have sought to investigate adolescent online safety across Africa, addressing key concerns such as cyberbullying, online harassment, digital literacy gaps, and privacy risks \cite{Hendricks2020Apr, ibrahim_vi, kunnuji2012online, tomczyk2021parents}. For example, using a non-participatory ethnographic study, Raochoene and Oyedemi \cite{Rachoene2015Jul} explored cyberbullying behaviors among South African teens, focusing on how they express themselves online. Their findings revealed that insults and threats, attacks on intelligence and physical appearance, sexting, and outing were among the most common cyberbullying behaviors. These results aligned with an earlier study by Ong'ong'a et al. \cite{ong2017utilization} on Kenyan teens aged 12–14, which also found a high prevalence of cyberbullying, with many victims experiencing repeated online harassment. While these studies provided insights into adolescent online risks, few have explored intervention strategies, such as parental mediation and educational initiatives tailored to the African context \cite{kritzinger2015enhancing, cyberbullet}.

Nigeria, as a multicultural nation with a complex sociocultural landscape, presents unique factors that shape adolescent online experiences. Socioeconomic inequality, digital literacy levels, cultural norms, and access to technology influence how young Nigerians experience cyberbullying, online harassment, and digital safety interventions. Despite the rapid digitalization of Nigerian society and increasing internet access among teens, empirical research on their online experiences remains scarce. Previous studies have examined specific risks, such as cyberbullying \cite{akeusola_cyberbullying}, online sexual risks \cite{kunnuji2012online}, and cybercrime among youth \cite{Adesina2022Sep}. \edit{Most} of this research focuses on \textit{"older adolescents}" and does not capture the full spectrum of online safety experiences among younger adolescents. For instance, Adesina et al. \cite{Adesina2022Sep} investigated cybercrime trends among Nigerian youth, identifying socioeconomic instability and poverty as key factors that drive engagement in cyber-related offenses. A 2021 study by Ibrahim and Vi \cite{ibrahim_vi} found that 7\% of surveyed secondary school teens had either engaged in or been exposed to cyberbullying, highlighting the widespread nature of the issue. While these studies provide valuable empirical insights, there has been no comprehensive research examining Nigerian adolescents' broader online experiences including their access to technology, exposure to risks, safety strategies, and the role of key stakeholders in shaping their digital interactions. By addressing these gaps, our study aims to contribute to the growing discourse on adolescent online safety in Africa, offering a holistic perspective on how Nigerian adolescents navigate online opportunities and risks in their socio-cultural and economic contexts.

\section{Method}
This section provides a comprehensive overview of the methodology employed in our study, including the research design and approach, data collection methods, study location, ethical considerations, and data analysis procedures. Each subsection outlines the steps taken to ensure the study's rigor, relevance, and ethical integrity.

\subsection{Research Design and Approach}
Our study aimed to explore the online experiences of secondary school adolescents in Nigeria. To provide context for the research, a preliminary literature review was conducted, examining publications from the past decade in the ACM Digital Library and AfriCHI conference proceedings. These sources were selected for their relevance to the SIGCHI community and their focus on online safety research. The review revealed a notable absence of studies that specifically addressed the online safety experiences of Nigerian adolescents, particularly those that engaged adolescent participants in the research process. To address this gap, we adopted a survey methodology to collect cross-sectional empirical data from a large sample. The target population comprised senior secondary school students, selected due to their developmental proximity to the minimum age requirements for social media use and their active engagement with digital technologies. \edit{Participants were recruited based on their enrollment in senior secondary school, with the age range usually between 12–18 years old.} Given the restrictions imposed by Nigerian school policies which typically prohibit the use of personal devices such as mobile phones during school hours, physical self-administered questionnaires were used instead of online surveys. This approach ensured accessibility for participants and compliance with institutional regulations.

\subsection{Data Collection}
This study was conducted in the Federal Capital Territory (FCT), Nigeria's capital city. The FCT is a melting pot of Nigeria's diverse population, representing people from all 36 states of the country. This diversity made it an ideal setting for obtaining a rich and representative sample for our study. A stratified simple random sampling approach \cite{acharya2013sampling} was employed to ensure balanced representation. Secondary schools within the FCT were categorized into strata based on their type: private and public schools. From these strata, 409 self-administered questionnaires \footnote{\href{https://osf.io/72e58?view_only=3e2700a6e1db46d99001842695536814}{Questionnaire Sample}} were randomly distributed across five schools (two private and three public) over the course of four weeks. \edit{Given a total secondary school population in the region of approximately 88,721 \cite{FCTSecondaryonline}, this sample size of (n=409) yielded a margin of error of ±4.83\% at a 95\% confidence level, assuming maximum variability. This indicates that the study sample provides a statistically reliable snapshot of adolescents’ online experiences within the FCT.}

\edit{In line with HCI’s emphasis on participant-centered methodologies \cite{mainstream_media, anuyah2023characterizing}, data collection was designed to be interactive and responsive to the needs of adolescents. All surveys were administered in English and completed by participants in their regular classroom settings, providing a familiar and comfortable environment that helped foster trust and openness. The first author was present at all survey sessions to support comprehension and ensure equitable participation. This in-person engagement allowed for real-time clarification and translation into local languages when needed, respecting linguistic and cultural diversity. These decisions were informed by HCI’s attention to the importance of minimizing barriers to participation, especially when working with youth populations across varying literacy and digital exposure levels.}

\subsection{Ethical Considerations}
This study adhered to rigorous ethical standards to ensure the protection, privacy, and informed participation of all respondents. Ethical approval was first obtained from the Federal Capital Territory (FCT) Secondary Education Board (SEB) \cite{Shehu2025Jan}, the official body overseeing public secondary schools in the region. Following this, we obtained formal permission from school principals and head teachers, who received detailed information about the study’s purpose, procedures, and participant protections. To comply with ethical standards for research involving minors, written informed consent was obtained from parents or legal guardians prior to participation. Consent forms outlining the study, data handling practices, and participant rights were distributed through the schools and returned signed. \edit{In addition to parental consent, assent was obtained from the adolescent participants themselves as they were informed both verbally and in writing about the voluntary nature of their participation, their right to withdraw at any point, and the measures taken to protect their anonymity}. All survey responses were completed anonymously, and no personally identifiable information was collected. These procedures were implemented to ensure participant confidentiality and minimize risk.

\subsection{Data Analysis}
This study employed a mixed methods design, integrating both quantitative and qualitative approaches to comprehensively address the research questions. A parallel design structure guided the analytical process, with each research question examined using the most appropriate methodological approach. For RQ1 and RQ2, we conducted quantitative analyses on closed-ended survey responses. Descriptive statistics were used to summarize the data: frequencies and percentages were computed for categorical variables such as device ownership, types of internet access, and exposure to online risks, while means and standard deviations were calculated for continuous variables where applicable. These analyses enabled the identification of broad patterns in adolescents’ access to digital technologies and the types of online risks they encountered. To address RQ3, we conducted a reflexive thematic analysis \cite{clarke2017thematic} of responses to a single open-ended survey question. The analysis followed a three-phase process: familiarization with the data, generation of initial codes, and development and refinement of themes. An inductive coding strategy allowed themes to emerge directly from participants’ responses, ensuring that the findings remained grounded in their lived experiences. By integrating quantitative and qualitative findings through methodological triangulation, this study enhances the depth, credibility, and explanatory power of the results.

\edit{\subsection{Theoretical Framework: Protection Motivation Theory}
To support the interpretation of findings related to RQ2 and RQ3, we employed Protection Motivation Theory (PMT) as an analytical lens. PMT provides a structured framework for understanding how individuals evaluate threats and decide whether and how to adopt protective behaviors in response \cite{tsai2016understanding, wilkinson2022many}. It is particularly well-suited for exploring how adolescents perceive online risks, interpret their severity, and determine appropriate safety responses within their digital environments. PMT outlines two primary cognitive processes that drive protective motivation: threat appraisal and coping appraisal. Threat appraisal refers to how individuals assess the risk itself. It includes perceived severity, the extent to which they believe the threat is serious or harmful (e.g., exposure to online predators or scams), and perceived vulnerability refers to how likely they think they are to personally experience that threat. In our context, this helped us understand how teens judged the dangers of harmful content or online interactions based on their lived experiences and awareness.
Coping appraisal focuses on how individuals evaluate their ability to deal with the threat. It includes response efficacy, or the belief that a particular protective behavior (such as blocking a user or reporting content) will be effective; self-efficacy, or the confidence that they can carry out that action successfully; and response costs, which are the perceived downsides or barriers to acting, such as fear of retaliation, lack of digital literacy, or parental misunderstanding. These dimensions helped us analyze not just whether teens recognized a risk, but also whether they felt capable and supported enough to respond to it. By applying PMT, our study interprets adolescent safety behaviors not as isolated decisions, but as situated responses shaped by both cognitive appraisals and the broader social, technological, and infrastructural contexts in which teens navigate online spaces.}

\section{Findings}
The findings from our study offer a comprehensive overview of Nigerian teens’ access to online technology, the online safety challenges they face, the strategies they employ to navigate these risks, and their recommendations for improving safety measures.

\subsection{Participant Demographics and Characteristics:}
\begin{figure}[h]
    \centering
    \begin{subfigure}[b]{0.40\textwidth}
        \centering
        \includegraphics[width=\textwidth]{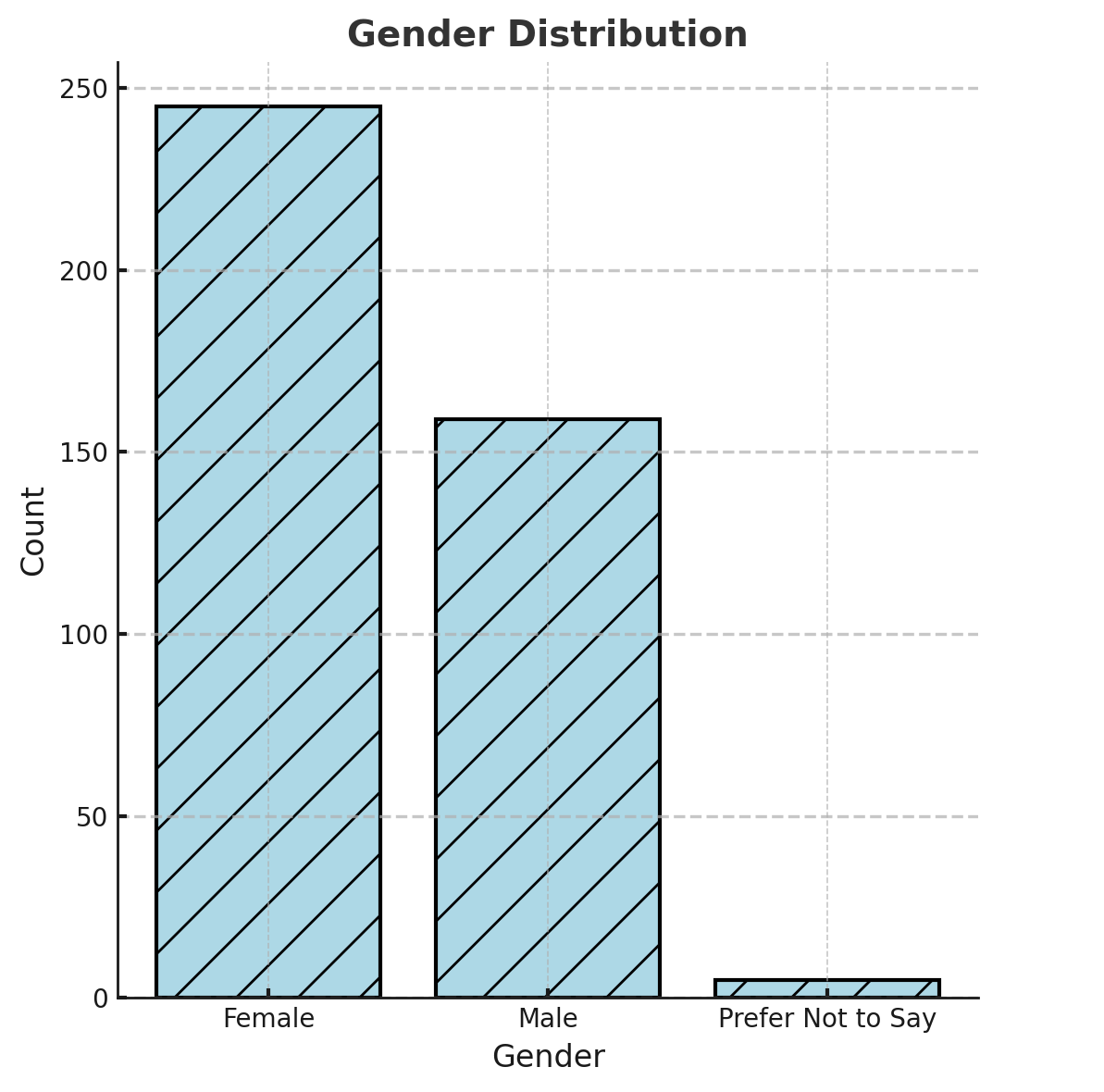}
        \caption{Gender distribution}
        \label{fig:gendr_dist}
    \end{subfigure}

    \vspace{2em} 
    \begin{subfigure}[b]{0.40\textwidth}
        \centering
        \includegraphics[width=\textwidth]{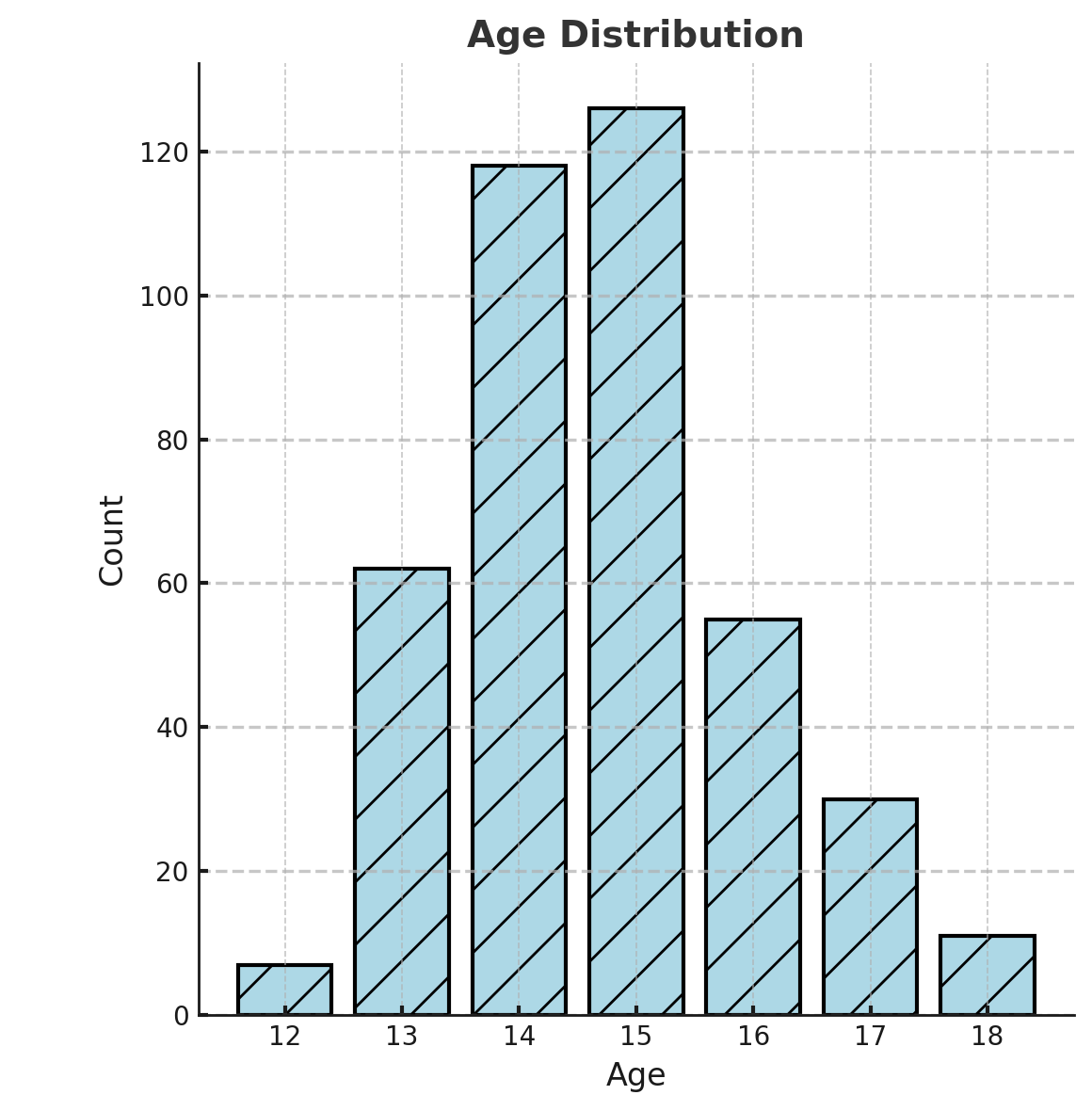}
        \caption{Age distribution of the sample}
        \label{fig:age_dist}
    \end{subfigure}

    \caption{Gender and Age Distribution of Participants}
    \label{fig:gender_age_distribution}
\end{figure}

Across the five schools surveyed, (N=409) respondents participated in our study, which reflected diverse gender identities and ages. A majority of participants identified as female, accounting for (60\%, N=245), while (39\%, N=159) self-identified as male. A small proportion (1\%, N=5) chose not to disclose their gender. Regarding age, participants in our study ranged from 12 to 18 years old  \textit{(M = 14.72, SD = 1.28)}, with the highest representation among 14-year-olds (29\%, N=118) and 15-year-olds (31\%, N=126). Adolescents aged 13 years accounted for (15\%, N=62) of our sample, while those aged 16 years made up (13\%, N=55). Smaller groups were represented by 17-year-olds (7\%, N=30), 18-year-olds (3\%, N=11), and 12-year-olds (2\%, N=7). This demographic distribution (See \autoref{fig:gendr_dist} and \autoref{fig:age_dist}) highlights a balanced mix of early and late adolescents, providing valuable insights into the perspectives of youth across a critical developmental period.

\subsection{Prevalence of Smartphone Use and Daily Internet Access (RQ1)}
To address RQ1, which explores the level and nature of access to online technologies among Nigerian secondary school adolescents, our findings revealed diverse patterns of device use, connectivity, and frequency of online engagement. 

\subsubsection{\textbf{Devices Used for Internet Access:}}
\begin{figure}[ht]
    \centering
    \begin{subfigure}[b]{0.40\textwidth}
        \centering
        \includegraphics[width=\textwidth]{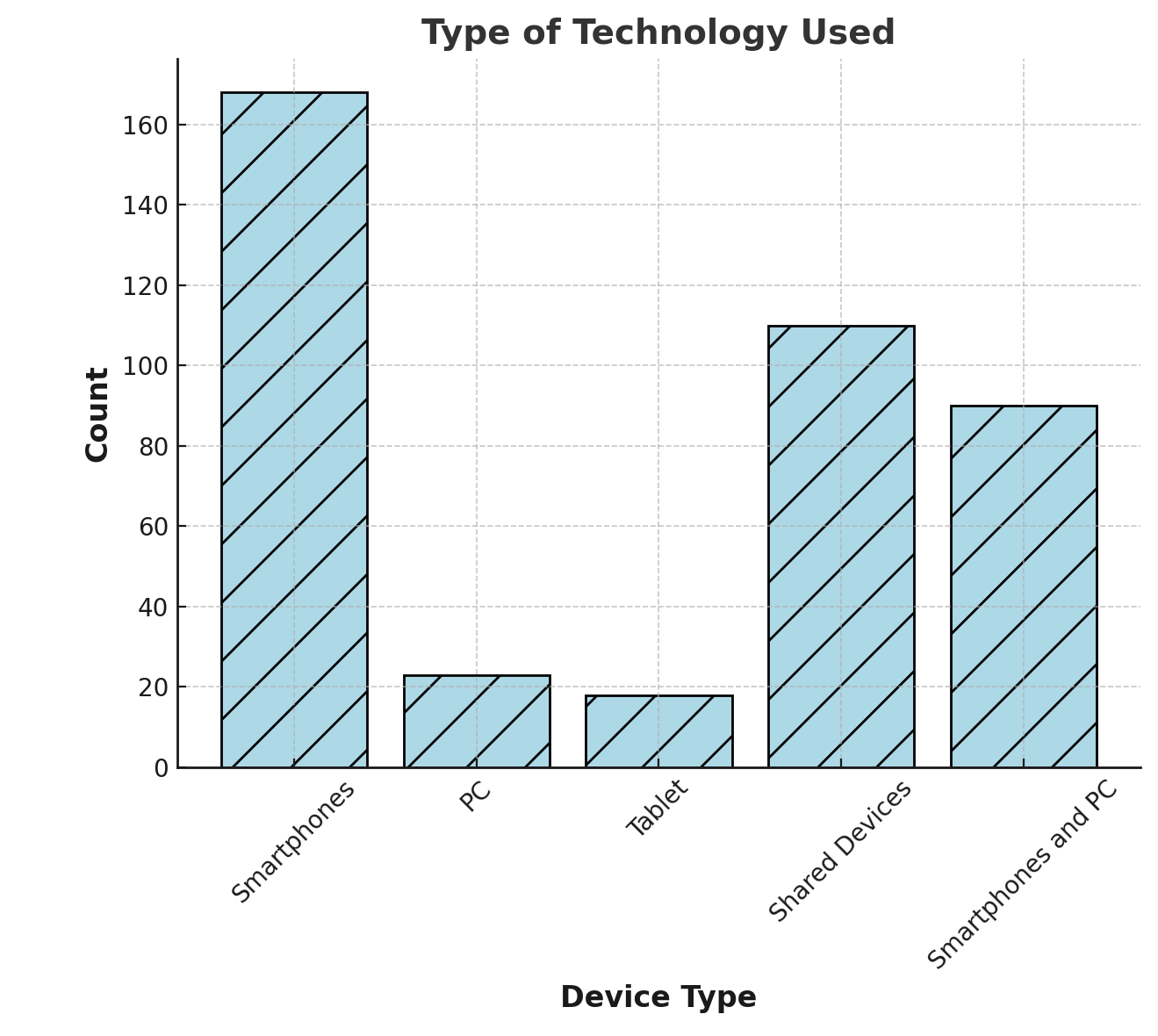}
        \caption{Common devices used to connect to the Internet}
        \label{fig:tech_use}
    \end{subfigure}

    \vspace{2em} 
    \begin{subfigure}[b]{0.40\textwidth}
        \centering
        \includegraphics[width=\textwidth]{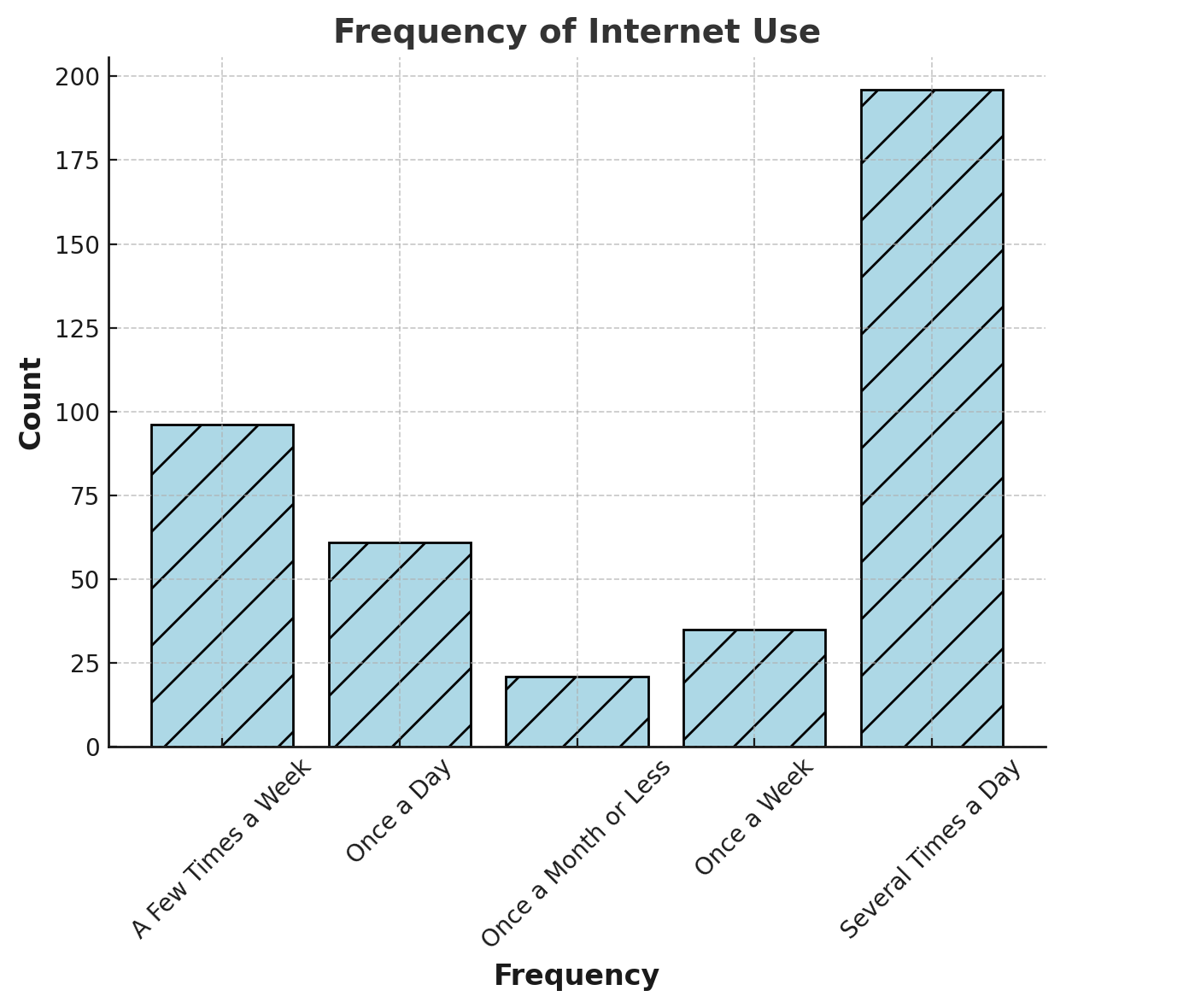}
        \caption{Frequency of Internet Use}
        \label{fig:int_use}
    \end{subfigure}

    \caption{Comparison of Internet Access and Technology Use.}
    \label{fig:access_tech_use}
\end{figure}

Majority of respondents (41\%, N=168) primarily used smartphones to access the internet, confirming their role as the most accessible and dominant digital tool among teens. Shared devices such as communal computers or family-owned gadgets were the second most reported mode of access (27\%, N=110), highlighting reliance on shared technological resources. A smaller portion of teens (22\%, N=90) used a combination of smartphones and personal computers, while exclusive use of personal computers (6\%, N=23) and tablets (4\%, N=18) was less common. These findings emphasize the centrality of smartphones in teens' digital lives, as well as the infrastructural limitations that lead many to depend on shared devices.
In terms of internet connectivity, mobile data was by far the most common mode of access, reported by 59\% (N=241) of respondents across both public and private schools. Home Wi-Fi followed at 19\% (N=79), and 17\% (N=70) reported having access to both home Wi-Fi and mobile data, indicating some flexibility in connectivity options. Access through school Wi-Fi (3\%, N=11) and public Wi-Fi (2\%, N=8) was minimal, reflecting limited institutional infrastructure for internet access (\autoref{fig:tech_use}). These trends underscore the critical role of mobile data as a gateway to the internet for many Nigerian adolescents.

\subsubsection{\textbf{ Frequency of Internet Use and Activity type:}} 
\begin{figure}[htb]
    \centering
    \includegraphics[width=0.50\textwidth]{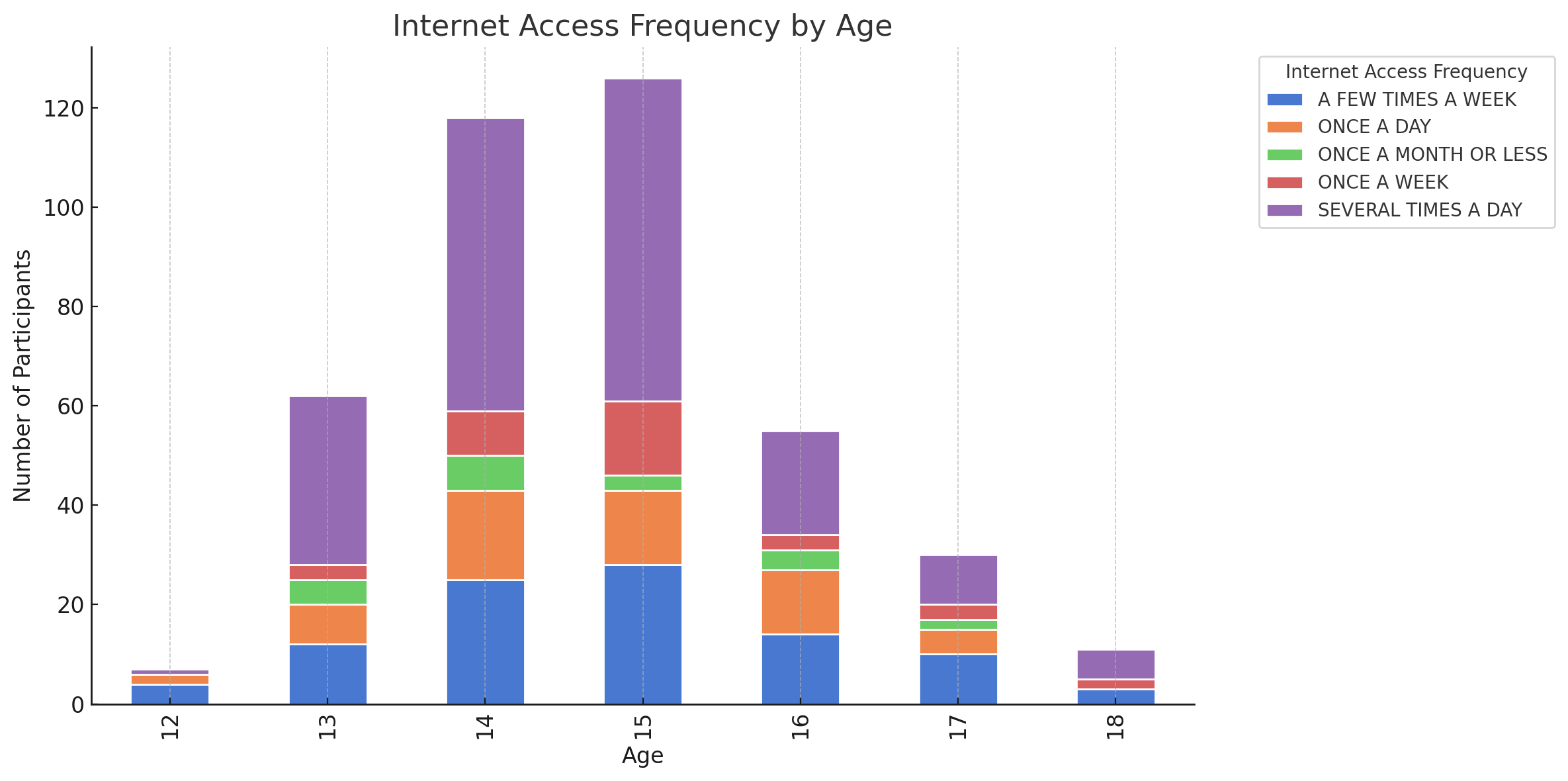}
    \caption{Age distribution of Internet Use}
    \label{fig:age_acsess}
\end{figure}

Nearly half of respondents (47\%, N=196) reported accessing the internet several times a day, indicating a high level of digital engagement. Others accessed the internet once daily (15\%, N=61) or a few times per week (24\%, N=96), while less frequent patterns were observed among those who used it once a week (9\%, N=35) or once a month or less (5\%, N=21) (\autoref{fig:int_use}). \edit{When disaggregated by age, the most frequent users were adolescents aged 14 and 15, who together accounted for nearly half of all respondents who accessed the internet several times a day (N=124). In contrast, younger participants (ages 12 and 13) and older teens (ages 17 and 18) reported less regular access, including weekly or monthly use (see \autoref{fig:age_acsess}). These trends suggest a developmental peak in digital engagement during mid-adolescence, which may have implications for their exposure to online risks and their familiarity with digital literacy tools.}

In terms of online activity, 60\% (N=244) of respondents used the internet primarily for social media, while 40\% (N=165) engaged in both social media and online gaming, highlighting the internet’s dual function as a space for both social interaction and entertainment. Taken together, these findings portray a digitally engaged youth population with widespread access to mobile technologies but also reveal disparities in access frequency and connectivity, shaped by infrastructural, developmental, and socioeconomic factors.

\subsection{Prevalence of Risks Posed by Inappropriate Online Content (RQ2)}
To explore the risky online experiences faced by secondary school students in Nigeria (RQ2a), we included a comprehensive list of potential online risks in our survey and asked participants to identify the ones they had encountered (See \autoref{fig:risk}). The analysis revealed that a significant majority of respondents, 67\% (N=272) reported no exposure to notable online risks. This suggests that a substantial portion of the surveyed adolescents either did not encounter harmful content or were unaware of such risks. \edit{From the perspective of PMT, this lack of perceived threat may reflect low perceived vulnerability or severity of risk, which can reduce the likelihood of engaging in protective behaviors. It is also possible that normalization of online risk, or limited digital literacy, contributed to underreporting/disclosure of risky online experiences.}
However, among those who reported experiencing risks online, 11\% (N=47) identified exposure to inappropriate content as the most common specific risk. This highlights the pervasive nature of unsuitable materials in the online environment and its accessibility to teens. The second most reported risk was online scams or fraud, experienced by 9\% (N=37) of respondents. This finding underscores vulnerabilities in teens’ financial literacy and awareness of deceptive online practices. Additionally, 5\% (N=20) of teens reported exposure to a combination of severe risks, including cyberbullying, identity theft, online sexual abuse or harassment, scams, and inappropriate content. This subset illustrates the compounded risks some teens face, potentially exacerbating the psychological and emotional toll of their online experiences. Less frequent but still critical were individual reports of specific risks such as identity theft (4\%, N=15), cyberbullying (2\%, N=9), and online sexual abuse or harassment (2\%, N=9). While these percentages are relatively small, their potential impact on victims’ mental health, social well-being, and sense of security makes them significant.

\subsection{Limited Online Safety Education and Reliance on Blocking and Reporting as a Safety Strategy(RQ2)}
Online safety education is critical in shaping adolescents’ coping appraisal by increasing both their awareness of risks and their confidence in managing them. To assess the availability and impact of such education (RQ2), our survey explored whether teens had received any formal or informal training on online safety and examined the strategies they commonly use to navigate or mitigate online risks.

\subsubsection{ \textbf{Online Safety Education:}}
The findings revealed a varied level of engagement with online safety education among the participants. Out of 409 respondents, 44\% (N=179) reported receiving some form of online safety education, while 28\% (N=115) stated they had not received any such education. Interestingly, an equal proportion 28\% (N=115) were unsure whether they had received online safety training. This uncertainty highlights a potential issue with how online safety education is delivered or communicated, suggesting either a lack of clarity, formalization, or access to these initiatives which could \edit{result in limited impact on students’ perceived self-efficacy}.
The relatively low percentage of teens explicitly reporting exposure to online safety education points to a significant gap in \edit{ adolescents' ability to appraise and respond to online threats. When education is absent or poorly communicated, teens may struggle to recognize risks (low threat appraisal) or feel uncertain about how to respond (low coping appraisal).} Furthermore, this finding underscores the need for more structured, accessible, and consistent programs to ensure that all teens are adequately equipped with the tools and knowledge to navigate the complexities of the online world safely.

\subsubsection{ \textbf{Online Safety Strategy:}}
\begin{figure}[ht]
    \centering
    \begin{subfigure}[b]{0.40\textwidth}
        \centering
        \includegraphics[width=\textwidth]{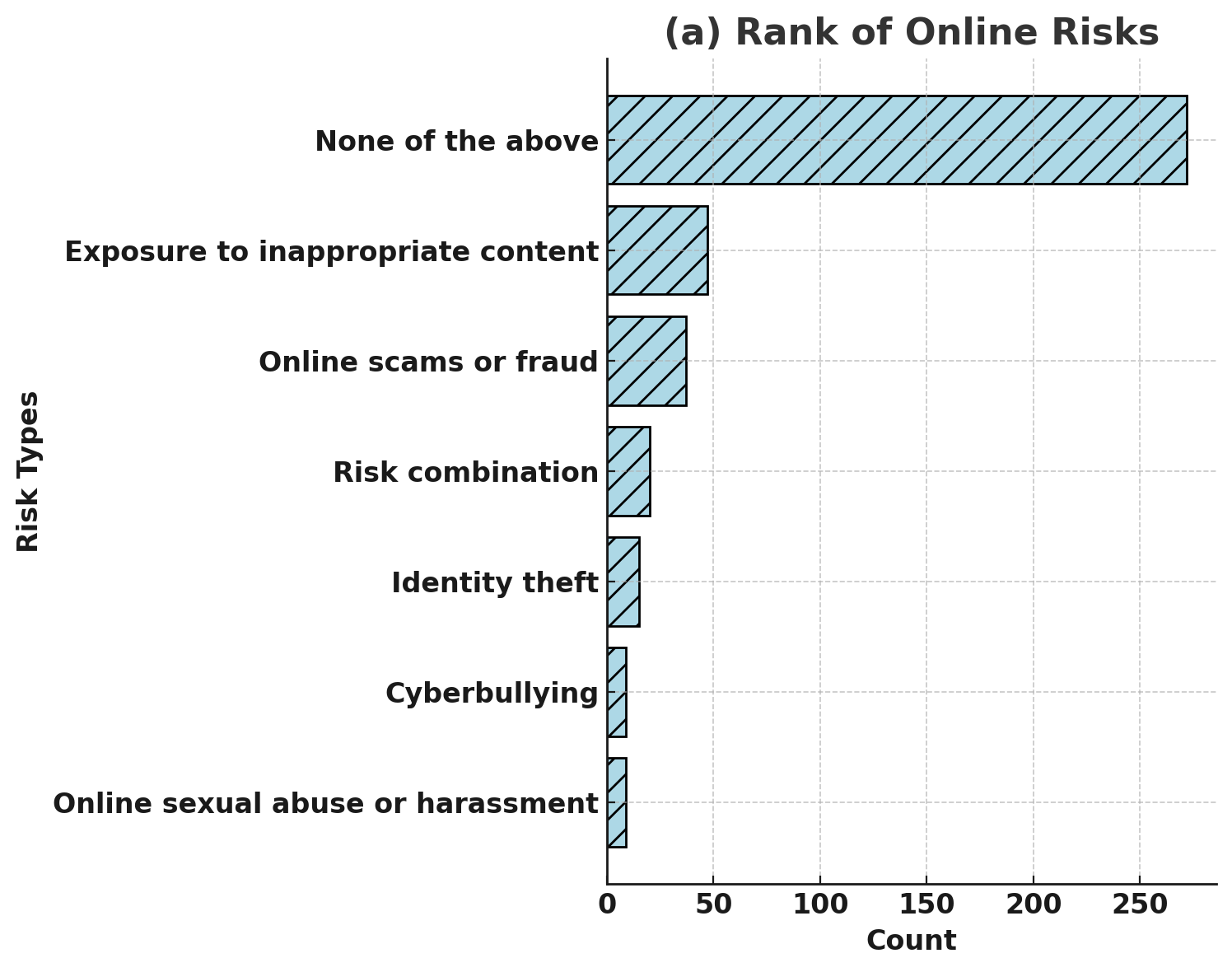}
        \caption{Ranking Online Risk Experiences}
        \label{fig:risk}
    \end{subfigure}

    \vspace{2em} 
    \begin{subfigure}[b]{0.40\textwidth}
        \centering
        \includegraphics[width=\textwidth]{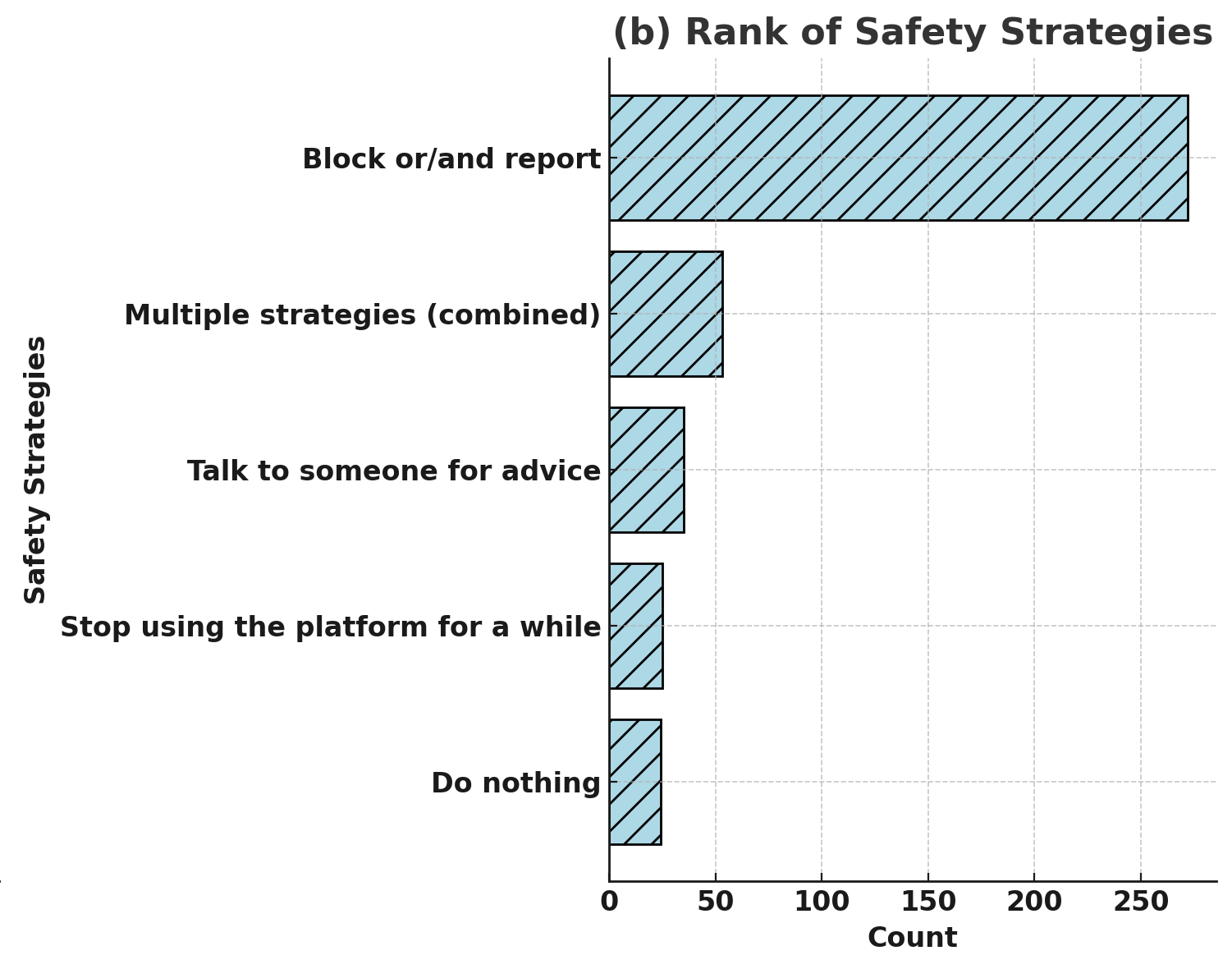}
        \caption{Safety Strategies Employed by Secondary School Teens in Nigeria}
        \label{fig:safety}
    \end{subfigure}

    \caption{Comparison of online risks and safety strategies.}
    \label{fig:risks_safety}
\end{figure}

When asked how they respond to risky online experiences (See \autoref{fig:safety}), the majority of respondents indicated relying on strategies that require minimal personal engagement, highlighting both strengths and limitations in their coping appraisals. Blocking or reporting the source of the risk was the most frequently reported strategy, used by 66\% (N=272) of respondents. This high rate suggests that these platform-provided tools are perceived as effective (high response efficacy) and easy to use (high self-efficacy), making them accessible first-line defenses against harmful content or interactions. The reliance on a single method may also reflect limited awareness of or confidence in alternative strategies. In addition to using platform tools, 13\% (N=53) of respondents reported employing a combination of strategies. These included blocking or reporting, temporarily discontinuing platform use, seeking advice from others, taking no action, or adopting alternative approaches, such as retaliating or completely closing the site. This finding suggests that some teens, particularly those facing more complex or severe online risks, preferred to diversify their strategies to mitigate harm. However, the relatively small proportion of teens combining strategies points to a potential gap in awareness or availability of these approaches.

Notably, just 9\% (N=35) of respondents reported seeking help from trusted individuals such as parents, teachers, or friends to navigate online risks. Although social support can be a highly effective form of protection, its limited use may reflect high perceived response costs such as teens worrying about being judged, misunderstood, or blamed, reducing their willingness to reach out. Furthermore, 6\% (N=25) of teens chose to temporarily stop using the platform as a way to avoid further exposure to harm. \edit{This strategy suggests high threat appraisal} and while practical for some, highlights the reactive nature of teens' responses to online risks and the potential for it to disrupt their engagement with digital spaces. Interestingly, 6\% (N=24) of respondents reported taking no action when confronted with online risks. This inaction could stem from various factors, including \edit{limited knowledge of available tools (low self-efficacy), a lack of confidence in their effectiveness (low response efficacy), or even the normalization of harmful online behaviors within their digital interactions (low threat appraisal)}. For some teens, the perception that online risks are unavoidable or insignificant may also contribute to their decision not to act.

\subsection{The Role of Parents in Supporting and Monitoring Teens' Online Interactions (RQ3)}
To understand the roles that parents play in managing adolescents' online safety, our findings underscore the pivotal role parents play in shaping teens' online experiences, both as a source of support and as active participants in monitoring and guiding online behavior. These insights highlight the importance of parental involvement in ensuring online safety while also revealing gaps and inconsistencies in the ways parents engage with their children's digital lives.

\subsubsection{ \textbf{Parents as the Primary Support Network for Teens:}}
When exploring the key stakeholders influencing Nigerian teens' online interactions, the findings revealed that parents were the most frequently consulted source of support. Among the 409 respondents, 36\% (N=149) reported turning to their parents when confronted with online risks or challenges. This reliance underscores the central role parents play in enhancing teens' coping appraisal by providing guidance and reassurance in navigating the digital landscape. Friends were the second-most common confidants, with 24\% (N=96) of teens seeking peer support. Interestingly, a notable portion of respondents (15\%, N=62) relied on a combination of support networks, including parents, friends, teachers, and siblings. This suggests that teens facing more complex or severe online issues are more likely to diversify their sources of support, seeking advice from multiple trusted individuals. However, 24\% (N=97) of participants reported not discussing their online experiences with anyone, signaling a significant gap in support for some teens. This lack of communication may stem from feelings of isolation, stigma, or the perception that no one in their network could adequately understand or assist them. Teachers, in particular, were rarely consulted, with only 1\% (N=5) of respondents turning to educators for help. This underutilization may reflect perceptions of the student-teacher relationship as less conducive to addressing personal or sensitive issues, as well as limited teacher involvement in online safety discussions \cite{mason2017conflict}.

\subsubsection{ \textbf{Parental Monitoring Practices and Their Inconsistencies:}}
Parental monitoring emerged as a critical factor in shaping teens' online safety. Among the respondents, 55\% (N=224) indicated that their parents actively monitored their online activities, while 29\% (N=117) reported no monitoring at all. Interestingly, 17\% (N=68) of teens were unsure whether their parents monitored their online behavior, highlighting a potential lack of explicit communication about monitoring practices within families. The frequency of monitoring also varied widely. While 27\% (N=111) of teens reported being monitored "often," the largest group 33\% (N=136) said monitoring occurred "sometimes." On the other hand, 23\% (N=93) described monitoring as "rarely," and 17\% (N=69) indicated it "never" occurred. These inconsistencies suggest that while a majority of parents are involved in their teens' online activities to some extent, the degree and consistency of their involvement vary significantly. This variability could be influenced by factors such as parents' digital literacy, work schedules, or differing views on the need for supervision.

\subsubsection{\textbf{Parents’ Understanding of Teens' Online Experiences:}}
The findings also shed light on teens’ perceptions of their parents’ understanding of their online experiences. While 54\% (N=220) of respondents felt that their parents "fully understood" their online activities, 12\% (N=49) believed their parents "only somewhat" understood, and 7\% (N=30) felt their parents did not understand their experiences at all. Additionally, 26\% (N=106) of teens were unsure whether their parents had any real understanding of their online activities, reflecting a lack of open communication in some families. A small proportion of participants (1\%, N=4) reported never being in a situation where parental involvement was required, indicating limited exposure to online challenges. These findings highlight the importance of fostering better communication between parents and teens about their online lives to bridge gaps in understanding and enhance the effectiveness of parental guidance.

These findings illustrate the crucial role parents play as stakeholders in shaping adolescent online safety, not only through supervision but also by providing support and mediating adolescents online interactions. However, gaps in understanding, inconsistent monitoring, and a lack of open dialogue suggest that this role is often underutilized or unevenly applied.

\subsection{Empowering Teens Through Online Safety Education and Awareness (RQ3)}
To further explore the role of adolescents in shaping their own online safety (RQ3), we concluded the survey with an open-ended question inviting participants to share their recommendations for improving the online safety of teens in Nigeria. Of the 409 respondents, 268 offered insights, resulting in the emergence of seven (7) key themes: (i) awareness and education, (ii) parental mediation, (iii) enhanced safety and privacy tools, (iv) stricter age restrictions, (v) improved content moderation, (vi) accountability and government regulation, and (vii) building resilience among teens. Below, we discuss each of these themes in more details:

\textbf{Awareness and Education:} A significant portion of respondents (N=103) underscored the critical need to raise awareness and educate adolescents about online safety. Participants highlighted the developmental changes teens undergo and how these can influence their self-regulation and decision-making, sometimes compromising their safety in favor of the perceived benefits of online technologies. To address these vulnerabilities, respondents advocated for comprehensive campaigns designed to equip teens with the knowledge and skills needed to navigate online risks effectively and practice safe behaviors. Suggested approaches included leveraging widely accessible channels such as social media platforms, schools, educational apps, mentorship programs, and community-based initiatives to educate and empower young users to make informed decisions and protect their privacy.
\begin{quote}
    \textit{"I feel like young people are sometimes ignorant of the vices online. \textbf{They need to educate and enlighten them about online safety}. They should know how to respond to online harassment and scams."} – P204 (14-year-old female)
\end{quote} 

\textbf{Parental Mediation:}
Many participants (N=51) emphasized the pivotal role of parental mediation in safeguarding teens from online risks. This perspective highlights the critical influence parents have in shaping safe digital behaviors and protecting their children from exposure to harmful content. According to some of the respondents, active parental mediation should involve a combination of active oversight, open communication, and providing consistent guidance on navigating online spaces safely. Respondents suggested that parental monitoring should extend beyond passive observation to include deliberate efforts to understand their children’s online activities. This could involve regularly reviewing content their children consume, checking for interactions that could pose risks (e.g., cyberbullying or scams), and intervening when necessary. 
\begin{quote}
    \textit{"Parental guidance should be advised for young children..."} – P301 (16-year-old female)
\end{quote}
In addition to engagement and communication, few teens also advocated for restrictive parental mediation, calling for stricter control over the devices and platforms they use. They recommended the implementation of technological tools, such as parental locks or parental control apps, to restrict access to inappropriate content or limit screen time. These tools, when combined with parental guidance, provide a layered approach to ensuring online safety.
\begin{quote}
    \textit{"I personally believe that every device a child owns \textbf{should have parental lock} on them."} – P375 (17-year-old female)
\end{quote}

\textbf{Enhanced Safety and Privacy Tools:}
A notable portion of respondents (N=42) highlighted the need to improve privacy and safety tools to enhance adolescents online security. Suggestions included developing user-friendly privacy features tailored to teens, such as tools to limit exposure to strangers, scam detection mechanisms, better reporting and blocking options, and localized educational guides on using these tools effectively.
\begin{quote}
    \textit{"Online safety for young people in Nigeria can be improved by the \textbf{creation of blocking and reporting tools for Nigerians by Nigerians} and by educating the public through social media and television on safety protocols..."} – P68 (15-year-old male)
\end{quote}

\textbf{Stricter Age Restrictions:}
Some respondents (N=32) advocated for stricter enforcement of age-appropriate access to online platforms. Recommendations included improved age verification mechanisms and measures to prevent younger users from accessing harmful content. Some participants suggested banning or limiting smartphone use for children below certain age thresholds to minimize their exposure to inappropriate material.
\begin{quote}
    \textit{"\textbf{Restrict the use of smartphones for children} under the age of 14, as it can lead to exposure to inappropriate content."} – P150 (14-year-old female)
\end{quote}

\textbf{Improved Content Moderation:}
Few participants (N=24) highlighted the importance of enhanced content moderation online to reduce exposure to harmful materials, including inappropriate or violent content and cyberbullying. Respondents suggested using advanced algorithms and human moderators to create safer online environments.
\begin{quote}
    \textit{"They should \textbf{block all pornography sites and ads} because they display rated pics and videos."} – P112 (18-year-old male)
\end{quote}

\textbf{Accountability and Government Regulation:}
A smaller but significant subset of participants (N=13) highlighted the importance of accountability and government regulation in ensuring a safer online environment for teens. They emphasized the need for stronger measures to hold individuals and organizations accountable for harmful behaviors or content online. This includes enforcing stricter penalties for offenses such as cyberbullying, fraud, and other forms of online misconduct. Participants suggested that dedicated agencies be established to address these issues, particularly those targeting adolescents.
\begin{quote}
    \textit{"There should be \textbf{agencies that fight against cybercrimes} so that the offenders can be arrested and punished."} – P293 (16-year-old female)
\end{quote}
The call for accountability reflects a perceived gap in current legal frameworks and enforcement mechanisms. Teens recognize that without tangible consequences for harmful actions, the prevalence of cybercrimes and online harassment may persist unchecked. Establishing specialized cybercrime units or agencies focused on online safety for adolescents could play a pivotal role in deterring harmful behaviors and creating a safer digital landscape.

\textbf{Building Resilience:}
A minority of the respondents (N=2) emphasized the importance of fostering resilience among teens by equipping them with critical thinking skills and coping mechanisms to handle online challenges effectively. Despite the reliance on parents to promote their safety, some teens acknowledge the need for empowerment interventions to help them identify and manage risky online situations.
\begin{quote}
    \textit{"I think everyone has to \textbf{learn to face online harassment.}"} – P10 (15-year-old male)
\end{quote}

Overall, this section presented various recommendations for promoting online safety among Nigerian teens, offering actionable insights for policymakers, educators, platform designers, and families. These suggestions emphasize the importance of collaboration across stakeholders to create a safer, more supportive, and empowering digital environment for adolescents.

\section{Discussion} 
This section explores the implications of our findings and provides tailored design recommendations to enhance online safety for Nigerian adolescents
\subsection{Digital Access, Online Risks, and Coping Strategies Among Nigerian Adolescents \edit{(RQ1 \& RQ2)}}
Our study underscores unequal access to the internet among Nigerian adolescents, with many facing barriers such as inconsistent connectivity, data costs, and limited device availability. These disparities contribute to a digital divide, where some adolescents are better equipped to engage with online resources safely, while others lack the necessary exposure, infrastructure, and digital literacy skills to navigate the online world effectively \cite{Livingstone2017}. These findings align with prior research on technology access gaps in low-income and developing nations \cite{wilkinson2022many, Quayyum2024Jun, Livingstone2017}, which highlight how economic constraints and infrastructural challenges determine who gets online, how often, and with what level of digital competence. Unlike Western contexts, where high connectivity is often available \cite{Livingstone2017}, our findings suggest that variations in digital access influence Nigerian adolescents’ exposure to online risks and their ability to seek support when confronted with online harm. Adolescents with frequent and unrestricted access may encounter more risks, but they also have greater exposure to online safety resources and protective strategies as revealed by western studies \cite{wisniewski_moral, zainab_nudges}. Conversely, those with limited or intermittent access may be less familiar with online risks, leaving them more vulnerable when they encounter harmful content or interactions.  

Additionally, our analysis reveals that some 12-year-olds reported accessing online technology despite the minimum age requirement for most social media platforms being 13 years old \cite{pew_age}. This suggests that many young users are engaging with online spaces before they are developmentally prepared, often without the necessary digital literacy or parental guidance to navigate online risks. Given that a significant number of younger users are already active online, this raises concerns about age verification enforcement on platforms as highlighted by earlier studies \cite{Odeigah2025Feb, marsden2023age} and the need for educational initiatives that equip preteens with foundational digital safety skills before they begin engaging in social media interactions. 

Finally, our findings revealed that one in three Nigerian adolescents encountered online risks. Among those who accessed the internet frequently, exposure to inappropriate content and online scams emerged as the most prevalent risks. However, a significant portion of adolescents reported not experiencing any online risks, raising questions about whether these risks are genuinely lower or simply normalized within their digital environments \cite{Ojanen2015Apr}. Prior research \cite{marwick2017nobody, Livingstone2017} suggests that adolescents in low-resource settings may downplay or internalize online harm, either due to limited awareness or the lack of available safety measures. When adolescents encountered online risks, blocking and reporting were their most frequently used coping strategies, reflecting a strong reliance on platform-provided tools, consistent with previous studies \cite{chatlani2023teen, alsoubai2022friends}. However, a notable proportion of adolescents took no action, signaling barriers such as fear of parental punishment, lack of trust in reporting systems, or the perception that taking action would not change their situation. This highlights the need for more robust and accessible online safety education that not only teaches adolescents to recognize risks but also provides effective, culturally relevant strategies to navigate them.

\subsection{The Role of Parents, Education, and Community in Online Safety \edit{(RQ3)}}
Parents emerged as the primary support system for Nigerian adolescents navigating online risks. However, our findings revealed inconsistencies in parental monitoring, with some adolescents experiencing strict control, while others were unsure whether their parents monitored them at all. This suggests that many Nigerian families lack open discussions about online safety, with parental mediation often taking the form of restrictive control rather than guidance and dialogue. This finding aligns with existing literature, which describes the tension between restrictive and active parental mediation \cite{peebles2024parental}. Prior research has shown that overly restrictive parental mediation can lead to conflict, as adolescents often perceive strict monitoring as an invasion of privacy or a limitation on their autonomy \cite{sharma2024parental, katie_resilienvce, Adigwe2020Aug}. In contrast, active parental mediation where parents engage in open discussions about online risks and empower teens to make informed decisions has been found to be more effective \cite{sharma2024parental}. Our study highlights the need for parent-focused digital literacy programs that equip guardians and caregivers with non-intrusive monitoring tools and strategies for fostering trust-based conversations about online safety with their adolescents \cite{kalmus2022towards}.

Beyond the home, formal online safety education remains scarce among Nigerian adolescents. A substantial proportion of respondents reported that they had either never received online safety education or were unsure if they had. This lack of structured education hinders adolescents' ability to recognize and mitigate risks \cite{luthfia2021role}, reinforcing the need for localized awareness initiatives that integrate online safety education into school curricula, community programs, and social media campaigns \cite{Chipangura2022Dec, dudek2025we}. Additionally, teachers and community figures played a minimal role in supporting adolescents’ online safety, highlighting an untapped opportunity for schools and local institutions to serve as digital safety hubs \cite{Chipangura2022Dec, sweigart2025takes}. Given that many adolescents feel uncomfortable discussing online risks with their parents, expanding safety education to schools, religious centers, and peer-led mentorship programs could provide more accessible and culturally relevant support structures \cite{sweigart2025takes}. By leveraging these existing community networks, online safety interventions can be made more inclusive, sustainable, and aligned with Nigerian adolescents' lived realities.

\subsection{Strengthening Platform Accountability and Policy Interventions \edit{(RQ3)}}
While adolescents recognized blocking and reporting tools as essential first-line defenses, concerns remain about the effectiveness of content moderation and platform accountability. Many participants noted that harmful content, including scams, cyberbullying, and explicit material, remained prevalent, suggesting that existing moderation policies may not adequately address region-specific online risks \cite{shahid2023decolonizing}. This raises critical questions about whether platform interventions are truly effective for Nigerian adolescents or whether gaps in moderation leave them vulnerable to digital harm. A major challenge is that most content moderation systems are designed with Western contexts in mind, often failing to capture localized digital threats, linguistic nuances, and cultural differences in how online harm manifests. For instance, slang, local languages, and context-specific online risks may not be effectively detected by global moderation algorithms \cite{shahid2023decolonizing}. Addressing these gaps requires platforms to develop regionally adapted safety measures that account for Nigeria’s unique digital environment. AI-powered content moderation that recognizes Nigerian languages and slang could improve the detection of harmful content, ensuring that harmful behaviors are flagged more effectively. Additionally, hiring local content moderators \cite{shahid2023decolonizing} who understand the cultural nuances of online interactions in Nigeria could strengthen moderation efforts. 

Beyond platform accountability, government regulation of online safety in Nigeria remains weak, with a general lack of trust in enforcement mechanisms \cite{policy_trust}. While Nigeria has existing cybersecurity and child protection laws \cite{Bello2025Jan, protection_bill}, many adolescents and parents are unaware of their rights or do not know how to report online crimes \cite{Femi-Oyewole2024Aug}. This reflects a broader challenge in African digital policy spaces, where legal frameworks exist but are rarely enforced effectively. Without clear implementation and enforcement, these policies fail to offer meaningful protections for adolescents navigating online risks \cite{Femi-Oyewole2024Aug}. Strengthening local enforcement mechanisms for online crimes such as cyberbullying, scams, and online harassment is crucial in ensuring that victims have accessible and responsive legal support. For policy interventions to be effective, governments must collaborate with technology companies, educators, and civil society organizations to create robust digital safety frameworks \cite{mainstream_media}. Clear communication of online safety policies is necessary, ensuring that digital rights and reporting channels are well publicized and easily accessible to both adolescents and their guardians. Holding platforms accountable \cite{social_account} for protecting young users is especially important in African contexts, where moderation policies tend to be less stringent compared to Western regions. Without external oversight, companies may deprioritize the safety needs of users in Nigeria and other African nations, further exacerbating the risks faced by adolescents online.

\subsection{Implications for Design}
Our study underscores the importance of developing interventions that are not only effective but also deeply attuned to the cultural and contextual realities of Nigerian adolescents. \edit{Drawing from our findings, we present the following targeted and actionable design implications}. These recommendations are grounded in the lived experiences of the adolescents we studied and are tailored to address their specific online safety needs within their social and infrastructural contexts:

\textbf{Localized Educational Campaigns and Tools:}
Awareness and education emerged as the most frequently suggested measures for enhancing online safety among Nigerian adolescents. To address this need effectively, online technology and social media platforms should prioritize localized educational campaigns that reflect the cultural and linguistic diversity of Nigerian adolescents \cite{asthana2017translation}. These campaigns should adopt simple, relatable illustrations of online risks and safety tips, making them accessible to adolescents. For instance, gamified and interactive features can be embedded into online platforms to periodically engage adolescents in online safety education \cite{cyberbullet, petrykina2021nudging}. Social media platforms can create short video content or ads in local Nigerian languages, and Pidgin English to provide practical tips, enabling adolescents to identify and mitigate risks \cite{anaemejeh2022sexual}. Such localized efforts ensure that the content resonates with its audience and fosters actionable learning.
Additionally, community-based programs can serve as a powerful reinforcement of these lessons \cite{akter2023takes}. Interactive workshops hosted by schools, religious institutions, and youth groups can leverage trusted local networks to promote online safety. Peer mentorship initiatives \cite{oguine2024internet}, where older teens guide younger ones in recognizing scams, avoiding harmful content, and responding to cyberbullying, provide relatable role models and foster a sense of community support. An example of such an intervention could be an app designed with gamified scenarios that reflect Nigerian realities that teach adolescents to identify phishing attempts disguised as fake advertisements (ads) or fraudulent online content. By combining engaging design with culturally relevant contexts, these tools can empower adolescents to make informed decisions about their online safety.

\textbf{Parental Mediation Tools:}
Our findings underscore the pivotal role parents play in ensuring adolescents' online safety, emphasizing the need for technology designers to create user-friendly tools that facilitate effective parental mediation \cite{Dedkova2023Dec, Ren2022Apr}, particularly in resource-constrained contexts. These tools must prioritize accessibility by being free or low-cost, making them affordable for families across diverse socio-economic backgrounds. For parents without smartphones, practical alternatives such as SMS-based alerts on their child’s online activities can enable meaningful involvement through basic mobile devices \cite{kumar2024smart}.
In addition to technological tools, platforms should integrate educational content tailored specifically for parents. Leveraging widely used communication channels like WhatsApp groups, community radio, and local marketplaces can ensure broad reach and engagement. These initiatives should also involve other influential adults in adolescents’ socio-ecological environments, such as educators, community leaders, and religious mentors, to equip them with the knowledge needed to proactively address online risks. A practical example could include features that allow parents to set app usage limits or restrict access to specific applications via SMS commands \cite{kumar2024smart}. This approach ensures that parents, regardless of their technological expertise or device capabilities, are empowered to support and guide their children effectively in navigating online risks.

\textbf{Better Age Verification and Platform Moderation:}
Many adolescents in our study expressed a desire for stricter age verification mechanisms to prevent underage exposure to harmful content. One potential solution involves leveraging national systems, such as Nigeria’s National Identification Number (NIN) database, to support more robust verification processes that align with local regulatory frameworks \cite{orunsoluni_NIN}. \edit{However, we recognize the significant privacy and ethical concerns associated with linking official personal data to social media accounts, particularly the risks of data misuse, surveillance, and children's limited understanding of digital consent. To mitigate these concerns, any integration with NIN should be accompanied by stringent data protection protocols, including third-party oversight, explicit opt-in parental consent, and restricted use solely for verification purposes. Alternatively, platforms could explore privacy-preserving verification methods, such as third-party age checks or digital tokens that confirm age without exposing sensitive information}. In addition to age verification, platforms can employ geolocation-based content moderation to restrict access to material flagged as inappropriate under Nigerian social and legal standards \cite{masud2024hate}. Harmful content, including pornography and violent material, underscores the need for moderation systems that are contextually aware and culturally sensitive. For younger users, a “youth-only” mode \cite{social4kids} could feature curated content and stricter filters to limit exposure to adult material or harmful trends \cite{masud2024hate}. AI-based moderation tools should be complemented by local human moderators familiar with Nigerian slang and sociocultural nuances, such as cyberbullying rooted in tribal or religious tensions. Proactive algorithms that flag or block explicit content prior to upload could also add an important layer of protection. Collaborations with Nigerian NGOs or youth-focused organizations could help train moderators and ensure that content moderation efforts are accurate, respectful, and locally relevant. These strategies collectively work toward safer digital experiences that are both effective and attuned to Nigerian adolescents’ realities.

\textbf{Building Resilience Through Cultural Narratives:}
Fostering resilience emerged as a long-term strategy to empower adolescents \cite{katie_resilienvce}, especially those from marginalized backgrounds. Resilience-building tools should integrate cultural storytelling that reflects Nigerian adolescents’ realities. For instance, animations featuring relatable characters navigating online challenges can teach critical thinking and coping skills in an engaging way. Incorporating religious and moral values through partnerships with faith-based organizations can further enhance these programs, resonating deeply with Nigerian societal norms. A Nollywood-style animated series aired on local television or YouTube \cite{Saba2024Feb} could weave actionable online safety tips into familiar narratives, making the lessons both relatable and memorable.

\subsection{Limitations and Future Research}
Several limitations of our work could inform future research directions. Firstly, during data collection, we encountered challenges in accessing adolescents from private schools due to stricter data policies compared to public schools. This limitation highlights the need for future research to include more diverse participant groups, ensuring representation across different educational and socioeconomic contexts.
Secondly, as the first study to adopt a teen-centric approach to understanding adolescents’ online safety in Nigeria, we employed a \edit{mixed-methods design}. While this approach was instrumental in identifying broad patterns,  it was limited in its ability to capture the nuanced and context-specific nature of the risks adolescents encounter, as well as the depth of their coping mechanisms and mediation strategies. To address these limitations, future research should incorporate more in-depth qualitative methods, such as interviews, scenario-based studies, and participatory design sessions, to explore these aspects more comprehensively. These methods would not only provide richer data but also empower young users to contribute to the design of tailored interventions that resonate with their lived experiences.
Furthermore, future studies could benefit from examining stakeholder perspectives around adolescents' broader social ecosystem, including parents, educators, policymakers, and platform designers. This holistic approach would provide a more comprehensive understanding of the online safety landscape and the interplay of various influences on adolescents' digital lives. Additionally, investigating the motivations behind adolescents’ use of online technologies and how they leverage these tools for support could yield insights \edit{on their online experiences}. Understanding these motivations could guide the development of more targeted and effective safety measures and interventions.
By addressing these limitations and expanding the scope of research, future work can contribute to a more nuanced and inclusive understanding of adolescents’ online safety, paving the way for evidence-based strategies that effectively protect and empower young users.

\section{Conclusion}
This study provides a comprehensive overview of the online safety challenges faced by Nigerian teens, highlighting key risks, strategies, and opportunities for improvement. Teens reported exposure to a range of online risks, including inappropriate content, scams, cyberbullying, and identity theft. Although many rely on platform tools such as blocking and reporting to address these risks, gaps in external support networks and limited access to online safety education remain significant challenges.
Parental involvement emerged as a critical factor in teens’ online safety, with many teens turning to their parents for guidance. However, inconsistencies in parental monitoring and a lack of tools tailored to the socio-economic realities of Nigerian families indicate a need for targeted interventions. Similarly, the findings underscore the importance of awareness campaigns, educational initiatives, and culturally relevant tools to empower both teens and parents in navigating online risks. The study also revealed actionable recommendations from teens, emphasizing the need for localized safety tools, stricter age restrictions, improved content moderation, and stronger accountability mechanisms. These insights highlight the importance of designing interventions that resonate with the cultural and technological contexts of Nigerian adolescents.



\bibliographystyle{ACM-Reference-Format}
\bibliography{Sections/References}

\newpage
\appendix


\end{document}